# Dual Effect of L-Cysteine on the Reorientation and Relaxation of $Fe_3O_4$-Decorated Graphene Oxide Liquid Crystals

**M. Zhezhu [1], Y. Melikyan [1], V. Hayrapetyan [1], A. Vasil'ev [1], H. Gharagulyan [1, 2]***

[1] A.B. Nalbandyan Institute of Chemical Physics NAS RA, Yerevan 0014, Armenia
[2] Institute of Physics, Yerevan State University, Yerevan 0025, Armenia

**Author to whom correspondence should be addressed:* herminegharagulyan@ysu.am

## Abstract

Recently, functional soft materials based on graphene oxide (GO) have attracted more attention due to their tunability and responsivity to external fields. From this point of view, exploring the magnetic effects on GO-dispersed colloidal liquid crystals (LCs) offers more opportunities to develop novel magnetically responsive systems. Specifically, controlling the orientational ordering of functionalized GOLC compounds is of great interest for both scientific purposes and technical applications. Here, we study the dynamics of $Fe_3O_4$-decorated L-Cysteine-functionalized GOLC director under an external magnetic field and analyze L-Cysteine's influence on the reorientation and relaxation time of the director. In particular, $Fe_3O_4$ nanoparticles were synthesized by solution-combustion method and added for altering orientational properties of GO which we synthesized electrochemically and functionalized by L-Cysteine. In addition, a comprehensive comparison of the director behaviour of GOLC and $Fe_3O_4$-decorated L-Cysteine-GOLC was undertaken to verify the tunability of the aforementioned systems. Furthermore, we demonstrate dual-effect of L-Cysteine on the magnetic field-induced alignment and relaxation time of GOLC systems, namely decrease in reorientation time and at the same time increase in relaxation time. Besides, micropattern creation and controlling in the drying drops of GOLC (net- and knit-like, flower-like, radial- and parallel-strip *etc.*) using a magnetic field were shown. The results of our studies could facilitate the fabrication of ordered and patterned tunable GOLC assemblies for a range of advanced applications.

**Abbreviations:** GO: graphene oxide; LC: liquid crystal; GOLC: graphene oxide liquid crystal; Cys: L-Cysteine; Cys-GOLC: L-Cysteine-functionalized graphene oxide liquid crystal; Cys-GOLC/$Fe_3O_4$: $Fe_3O_4$-decorated L-Cysteine-functionalized graphene oxide liquid crystal; MNP: magnetic nanoparticle; POM: Polarized Optical Microscopy; DLS: Dynamic Light Scattering; UV-Vis Spectroscopy: Ultraviolet-Visible Spectroscopy; XRD analysis: X-ray Diffraction analysis; FTIR-ATR Spectroscopy: Fourier Transform Infrared - Attenuated Total Reflectance Spectroscopy; HR XPS: High-Resolution X-ray Photoelectron Spectroscopy; SEM: Scanning Electron Microscopy; EPR: Electron Paramagnetic Resonance; VSM: Vibrating Sample Magnetometer.

## Introduction

For the last decade, our understanding of graphene oxide (GO) has continuously evolved and improved largely due to its similarity to graphene in its reduced form [1, 2]. GO endowed with excellent physicochemical properties, in particular, it stands out with its biocompatibility, water dispersibility, fluorescence-quenching capability, nonlinear optical properties, *etc*., due to the enriched surface functional groups providing ideal interfaces and active sites [3, 4]. Meanwhile, incorporating diverse nanoparticles into the GO or its functionalization with different organic molecules, such as amino acids present notable improvements in the original characteristics of GO or the extension of new functions to the GO's properties [5-7].

From this perspective, L-Cysteine (Cys) plays a key role in modification of GO due to its reducing and functional properties [8, 9]. Reduction and functionalization of GO with Cys has quite interesting applications, such as it can be utilized for detecting environmentally harmful metal ions [10], improvement of anti-corrosion properties [11], multi component synthesis as recoverable catalyst [12], sensing in nanoarchitectonics [13, 14], *etc*.

The field of applications of GO is expanded not only due to its functionalization, but also by the formation of its liquid crystalline (LC) phase [15, 16]. LC phase of GO exhibits a significant responsiveness to external stimuli [17], enabling its use across a broad spectrum of applications [18]. It is even more interesting when these two advantages, namely GO functionalization and its LC phase formation are combined, which is presented in [19]. The orientational order of GO flakes can be controlled by its LC phase, which is extremely advantageous for improving the properties of these anisotropic materials [20]. Controlling LC alignment along a preferred direction is crucial for the fabrication of different devices [21], such as LC displays, modulators and light shutters [22]. The ability to manipulate micrometer-sized particles plays an important role in sensing and biological applications [23].

Another option for expanding the field of application of GO is the incorporation of magnetic nanoparticles (MNPs), namely $Fe_3O_4$ into it which offers a versatile and multifunctional platform for potential applications due to several synergetic properties [24-26] On the other hand, L-Cysteine-functionalized MNPs are also highly intriguing due to the opportunities they offer in terms of applications [27]. The capability to control and manipulate these composites with external magnetic fields makes them especially appealing for innovative research and technological development. It is possible to achieve color switching in a magnetically responsive photonic crystal of GO nanosheets by changing the direction of applied magnetic fields [28, 29]. Magnetic fields are used to induce stable LC ordering [30], however, magnetic field-induced orientation relaxes after the field is switched off that's why various methods are being developed to overcome this difficulty [31]. GOLC alignment can be controlled without using external electric and magnetic fields as well, namely it can be achieved based on the appropriate directional design of the solid-liquid and air-liquid interfaces throughout the rectangular and circular capillary tubes [32].

Patterned GOLC in drying drops offers an innovative approach to build multifunctional-performance optoelectronic devices with unique structures and tailored conditions [33]. Since Cys acts as a reduced agent, the patterned GOLCs can offer electrical conductivity due to the precise control over the material's structure at the microscale, additionally the anisotropy harnessed to create materials with controlled optical, electrical, and mechanical characteristics, which is essential for optoelectronic devices. Moreover, GO exhibits properties characteristic to those of photonic crystals, and patterned colloidal photonic crystals have an extremely broad range of applications [34]. All of this indicate that research in this area is highly relevant.

This work is devoted to the investigation of orientational behaviour of Cys-GOLC system under an external magnetic field with the presence of $Fe_3O_4$ nanoparticles in the structure. The director configurations and their transformations under the applied magnetic field have been studied. Additionally, relaxation dynamics is also studied. Besides, micropatterning of GOLC in drying drops and its tuning possibilities with the applied magnetic field are also explored.

## Materials and Methods

**Materials.** The graphite>99 % (CAS No.: 7782-42-5), L-Cysteine>99 % (CAS No.: 52-90-4), Nitric acid 70 % (CAS No.: 7697-37-2), Fe powder >99 % (CAS No.: 7439-89-6), glycine >99 % (CAS No.: 56-40-6) used in the experiments were purchased from 2D Semiconductors and Sigma-Aldrich Chemical Co.

### 1.1. Preparation of $Fe_3O_4$-decorated Cys-functionalized GOLC

GO synthesis, its LC phase formation and Cys-functionalization are well described in our previous works [13, 19]. For the next step, at first, $Fe_3O_4$ nanoparticles were synthesized using the conventional solution-combustion technique. 3 g of Fe metal (99,99%) was dissolved in dilute $HNO_3$ (30 ml) in a conic flask. Then 20 ml of deionized water was added. Following the fuel, glycine (10 g) was dropped and dissolved in the same flask. The brownish transparent solution was evaporated into a viscous gel which was spontaneously ignited. The resulting powder was grounded in an agate mortar and was calcined at 800$^o$C for 3 h in an air atmosphere. The mean diameter of synthesized nanoparticles was 600 nm. Actually, our $Fe_3O_4$ NPs, with 600 nanometers in diameter, are in fact agglomerates of smaller NPs (≤ 200 nm), which are bound together due to ferromagnetic interaction. The average crystallite size (25–30 nm) of the $Fe_3O_4$ NPs was estimated using powder diffraction pattern analysis in Rigaku SmartLab Studio 2 software, employing the Halder-Wagner method. The ferrolyotropic LC composites were prepared by precisely doping and stirring the Cys-functionalized GOLC with the $Fe_3O_4$ nanopowder for 30 minutes. The concentration of $Fe_3O_4$ nanoparticles in the composite was 1%. In Fig. 1 schematic representation of the chemical procedure is presented for the synthesis of $Fe_3O_4$ nanoparticles (Fig1(a)) and step-by-step fabrication of Cys-GOLC/ $Fe_3O_4$, including Cys functionalization of GOLC followed by decoration with the synthesized $Fe_3O_4$ MNPs.

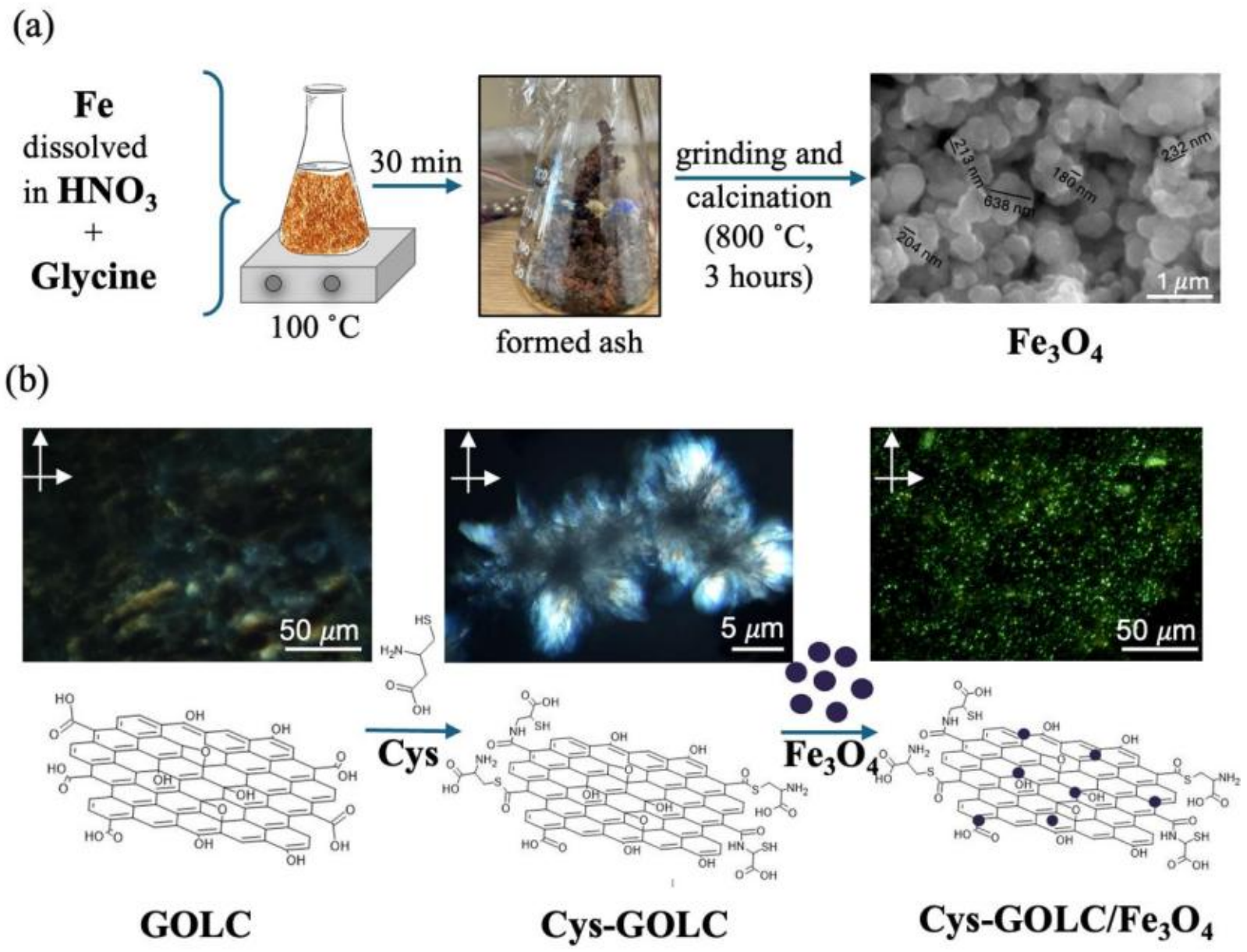

Fig.1 Schematic representation of the chemical procedure: (a) synthesis of $Fe_3O_4$ nanoparticles using the conventional solution-combustion technique and (b) step-by-step fabrication of Cys-GOLC/ $Fe_3O_4$, including Cys functionalization of GOLC followed by decoration with the synthesized $Fe_3O_4$ MNPs via an ultrasound-assisted process. Notably, for the preparation of GOLC/ $Fe_3O_4$, the Cys functionalization step was excluded.

### 1.2. $Fe_3O_4$-decorated Cys-functionalized GOLC cell preparation

For the experiments, sandwich cells were made using pre-cleaned microscope glass substrates without treatment and adjusted with 20 µm polyethylene spacers. Three types of materials, namely GOLC, $Fe_3O_4$-decorated GOLC and $Fe_3O_4$-decorated L-Cysteine-functionalized GOLC were taken and drop-filled for the cell preparation. Finally, the cells were glued with UV curable glue and studied by POM, DLS and UV-Vis Spectroscopy under the applied magnetic field using NdFeB magnets of 10×10×40 mm sizes. The magnetic flux intensity was 0.3 T. Additionally, a drop-drying approach was utilized for the GOLC patterning.

### 1.3. Characterization

The GO, $Fe_3O_4$-decorated GO, Cys-GO and $Fe_3O_4$-decorated Cys-GO systems were characterized using various microscopic and spectroscopic techniques. Particularly, crystallographic data for these systems were obtained through XRD analysis (Rigaku MiniFlex instrument with Cu*Kα*-radiation and scanning range: -3-145$^0$(2θ)). The chemical composition of the materials was analyzed with FTIR-ATR spectroscopy (Spectrum Two device, PerkinElmer; Spectral range: 4000 to 400 cm$^{-1}$). Bond types and hybridization of the above-mentioned systems were identified via Raman spectroscopy (LabRAM HR Evolution, HORIBA; Wavelengths range: 200-1100 nm, Light source: 633 nm, with a 600 gr/mm grating). The chemical composition of $Fe_3O_4$-decorated Cys-GO was determined through HR XPS analysis (KRATOS Axis Supra+ instrument, Shimadzu). Optical properties of the materials were examined using UV-Vis spectrophotometry (Cary 60 instrument, Agilent, Spectral range: 190 - 1100 nm). The morphology of the structures was analyzed using SEM (Prisma E, Thermo Fisher Scientific). Paramagnetic properties of GO were characterized by EPR spectrometer (built in Semenov Institute of Chemical Physics). Ferromagnetic properties of the synthesized $Fe_3O_4$ nanoparticles were studied by vibrating sample magnetometer (VSM-220, Xiamen Dexing Magnet Tech. Co., Ltd., China). Particle size analysis and zeta potential's value of the synthesized and functionalized materials were implemented utilizing the DLS technique (Litesizer 500, Anton Paar; Light source: 658 nm, 40 mW). LC phase of GO and above-mentioned composites were investigated by POM (MP920, BW Optics).

## 2. Results and Discussion

### 2.1. Structural and spectral analysis of $Fe_3O_4$-decorated Cys-functionalized GO

Characterization on the GO and Cys-GO is thoroughly detailed and described in our previous study [19]. Additional microscopic pictures and characteristic spectra of the GO, Cys-GO, as well as characteristics on the $Fe_3O_4$ and Cys-GO/$Fe_3O_4$ system are presented in Fig.2. The magnetic properties of the latter were explored at room temperature through the VSM technique in the range of -15000 to 15000 Oe. Based on the magnetic hysteresis curve, the magnetization is saturated at $M_s$=63.4 emu/g, while $M_r$=8.3 emu/g (remnant magnetization), and $H_c$=86.8 Oe (magnetic coercivity) (Fig. 2(a)) [35]. VSM measurements have done not only for $Fe_3O_4$ MNPs, but also for Cys-GO, GO/$Fe_3O_4$ and Cys-GO/$Fe_3O_4$ systems. It was revealed that Cys-GO did not exhibit any ferromagnetic response as expected. While GO/$Fe_3O_4$ and Cys-GO/ $Fe_3O_4$ show ferromagnetic response clearly [36, 37] (see in Supplementary Materials Fig. S1(a)). Particularly, the saturation magnetization of GO/ $Fe_3O_4$ and Cys-GO/ $Fe_3O_4$ is 2.1 emu/g and 1.4 emu/g, respectively, much smaller than that of $Fe_3O_4$ (63.4 emu/g), confirming the presence of magnetic $Fe_3O_4$ nanoparticles in these systems. It is reasonable that the $M_s$ values of the $Fe_3O_4$ - decorated systems are significantly lower than those of the pristine $Fe_3O_4$ nanoparticles, as the GO and Cys-GO are paramagnetic which is shown in Fig. 2 (b). Paramagnetic properties of GO and Cys-GO were also studied at the room temperature (Fig. 2(b)). EPR spectrum of the powdered GO is a narrow signal indicating a uniform distribution of paramagnetic centers [38, 39]. On the other hand, a narrow EPR linewidth suggests weak spin-spin interactions and weaker interactions minimize random distortions in the flake alignment, supporting the development of a long-range orientational order typical of LC phases. The broader linewidth in Cys-GO may reflect enhanced interactions or more complex dynamic behavior. For the Cys-GO the signal is more intense compared to that of GO. This could be as a result of the increased magnetic anisotropy and spin-coupling effects. Fig. 2(c) shows SEM image of Cys-GO/$Fe_3O_4$ system. As it can be seen, in certain areas of the sample, a very thin and nearly transparent layer of GO covers the $Fe_3O_4$ nanoparticles, with Cys sheets clearly visible beneath [13]. Fig. 2(d) depicts the Raman point-by-point 2D mapping, where the systems of our research, namely GO, Cys-GO, $Fe_3O_4$ and Cys-GO/$Fe_3O_4$ are clearly indicated. Raman spectra of Cys-GO and Cys-GO/$Fe_3O_4$ composite are presented in Fig. 2(e). Cys-GO shows the presence of two peaks, namely D-band at 1329.4 $cm^{-1}$ which corresponds $sp^3$ defect states and G-band found at 1598.1 $cm^{-1}$ which ascribed vibration of $sp^2$ hybridized C atoms of GO. The 2D-band is observed at 2650.7 $cm^{-1}$ which provides important insights on the stacking order of GO sheets, moreover the presence of D+G peak at 2924.5 $cm^{-1}$ is due to the resonance of D and G phonon modes [40]. Cys exhibits peaks at 198 $cm^{-1}$, 497 $cm^{-1}$, 676 $cm^{-1}$, 784 $cm^{-1}$, 2917 $cm^{-1}$, and 2968 $cm^{-1}$ corresponding to the CCS, COO, CS, $CH_2$ rocking mode, $CH_2$ symmetric and asymmetric stretching vibrations, respectively [41]. The $I_D/I_G$ ratio for this system is 1.1. Accordingly, the Raman spectrum of Cys-GO/$Fe_3O_4$ composite exhibits the D peak at 1328.3 $cm^{-1}$, and the G peak at 1586.6 $cm^{-1}$. The 2D peak at 2644.3 $cm^{-1}$ and the D+G peak at 2929.8 $cm^{-1}$ are also clearly expressed. The $I_D/I_G$ ratio in this case is 1.2. Incorporation of $Fe_3O_4$ to Cys-GO system reveals two distinct peaks $A_{1g}$ and $E_g$, corresponding to the symmetric stretching modes of oxygen atoms along the Fe-O bonds at 688 $cm^{-1}$ and vibration mode at 321 $cm^{-1}$, respectively [42]. $T_{2g}$ vibration mode of $Fe_3O_4$ at 549 $cm^{-1}$ is weakly expressed. Cys exhibits peaks at 201 $cm^{-1}$, 497 $cm^{-1}$, 683 $cm^{-1}$, 784 $cm^{-1}$, 2916 $cm^{-1}$, 2968 $cm^{-1}$. When comparing Cys-GO with Cys-GO/$Fe_3O_4$, it is clear that first two peaks differ in their stretching others distinguish by their intensities and slightly by positions. Fig. 2(f) represents typical FTIR-ATR spectra of Cys-GO and Cys-GO/$Fe_3O_4$ composites. As observed, the stretching of C-C, S-S, C-S, OH (out of plane band), $NH_3$ (asymmetric), C-C, C-O-C, S-H, C-OH, C-O-C, C-OH, C-N, CO-H, $NH_3$ (symmetric), C-OH, $COO^-$ symmetric, NH, $COO^-$ (asymmetric), C=C, C=O bands at 537 $cm^{-1}$, 614 $cm^{-1}$, 673 $cm^{-1}$, 711 $cm^{-1}$, 774 $cm^{-1}$, 845 $cm^{-1}$, 873 $cm^{-1}$, 964 $cm^{-1}$, 1041 $cm^{-1}$, 1088 $cm^{-1}$, 1124 $cm^{-1}$, 1193 $cm^{-1}$, 1296 $cm^{-1}$, 1335 $cm^{-1}$, 1381 $cm^{-1}$, 1403 $cm^{-1}$, 1480 $cm^{-1}$, 1579 $cm^{-1}$, 1621 $cm^{-1}$, 1658 $cm^{-1}$ are observed respectively [8, 43, 44]. However, for the Cys-GO/$Fe_3O_4$ composite the peak at 566 $cm^{-1}$ observes, attributed to the Fe-O vibrational mode of $Fe_3O_4$ [42, 36]. XRD spectra of the GO, Cys-GO and Cys-GO/$Fe_3O_4$ are shown in Fig. 2(g). In comparison with GO which has a diffraction peak at 8.4° (meaning a larger interlayer distance), Cys has peaks at 8.1°, 16.3°, 24.5°, 32.9° and $Fe_3O_4$ has a maximum at 29.9° (220) and at 33.7° (311) [13, 42, PDF-2 2022]. The analysis demonstrated the presence of reflections from the monoclinic particles of Cys, which are strictly oriented along the crystallographic c-axis, as evidenced by the presence of the intense peaks of (l00) predominantly from a single plane. This indicates a clear anisotropy of the material. The peak at 41.5° suggests reflections from rGO, formed as a result of interaction with Cys molecules, which is consistent with literature data [45] (see in Supplementary Materials Table S1). The atomic percentages for the Cys-GO/$Fe_3O_4$ are: C 44 % at %; O 32.7 % at %; N 5 % at %; S 16.1 % at %; Fe 1.1 % at %; Si (substrate) 1.1 % at %, respectively (see Fig. S1(b) in Supplementary Materials). The zeta-potentials of the synthesized GO, Cys-GO, GO/$Fe_3O_4$ and Cys-GO/$Fe_3O_4$ at pH=7 is shown in Supplementary Materials (see Fig. S1(c)). The positive charge of $Fe_3O_4$ MNPs is consistent with the formation of $FeOH_2^+$ in a basic environment [46], while the negative charge of Cys-GO is attributed with the presence of negatively charged functional groups on GO. The zeta potentials of the aqueous GO, GO/$Fe_3O_4$, and Cys-GO/$Fe_3O_4$ systems studied in this research are all negative and even more negative than -30 mV, indicating that these systems form stable suspensions due to interparticle electrostatic repulsion [47]. Besides stability, it plays an important role in determining orientation and interactions of particles in colloidal systems generally.

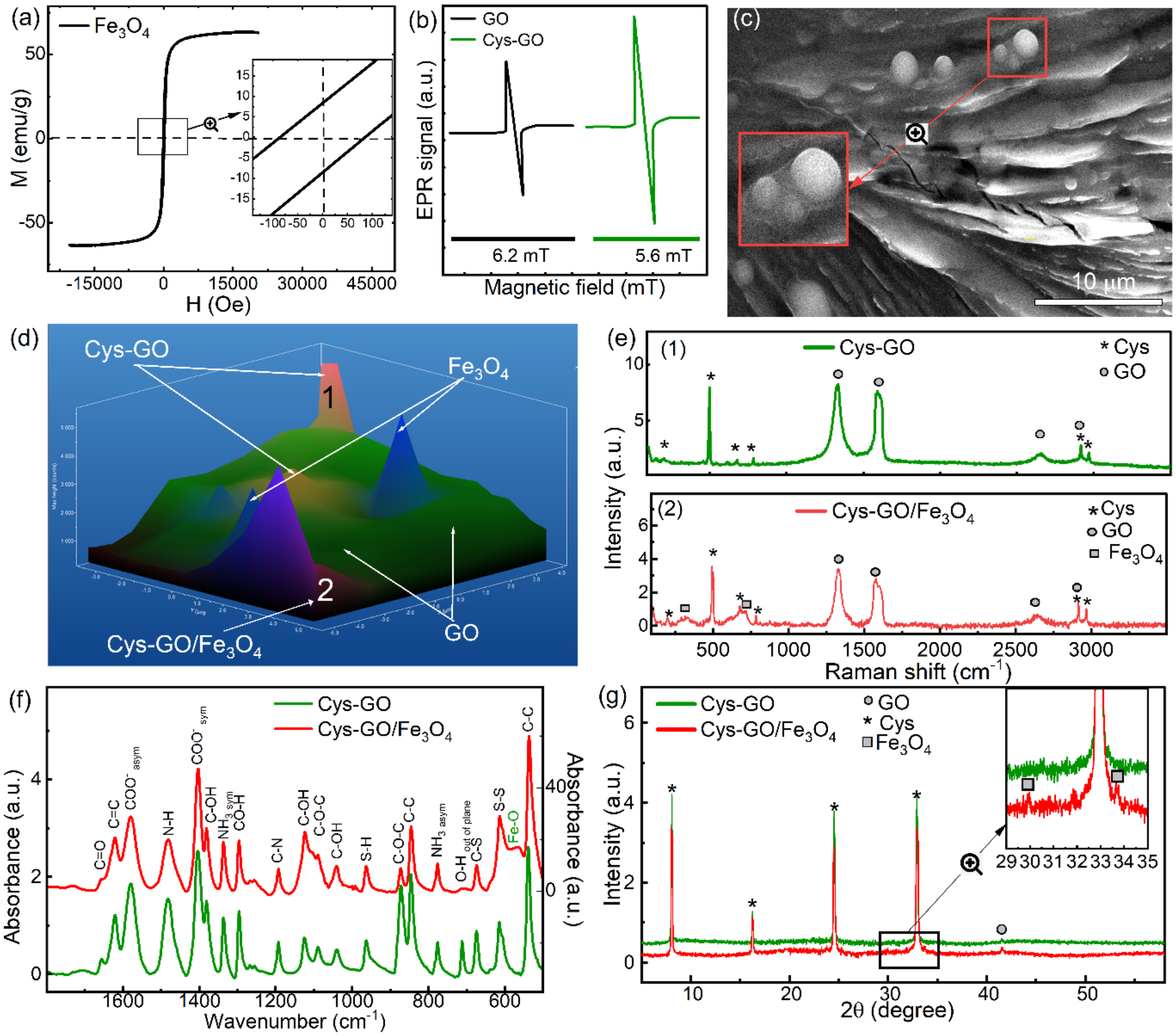

Fig. 2. Characterization of the synthesized, functionalized and decorated materials: (a) Magnetic hysteresis loop of $Fe_3O_4$ nanoparticles. (b) Room temperature EPR spectra of GO and Cys-GO at 9.468 GHz with 100 kHz modulation. $Mn^{2+}$ ion served as a standard. (c) SEM image of Cys-GO/ $Fe_3O_4$. (d) Raman mapping of the Cys-GO/$Fe_3O_4$ structure for the following parameters: Filter: 5 %; Acquisition time: 10 s (e) Raman spectra of Cys-GO and Cys-GO/ $Fe_3O_4$ composite. (f) FTIR-ATR spectra of Cys-GO and Cys-GO/ $Fe_3O_4$ composite. (g) XRD spectra of GO, Cys-GO and Cys-GO/ $Fe_3O_4$. All measurements were performed at a temperature of 25°C.

XPS studies of GO and Cys-GO are presented in our previous work [13]. Here we present the XPS analysis of Cys-GO/$Fe_3O_4$ composite conducted by the HR XPS Axis Supra+, with spectral analysis performed using CasaXPS software. Initially, a rapid screening was conducted to obtain a broad-spectrum scan of the XPS spectrum for the Cys-GO/$Fe_3O_4$ (see Fig. 3(a)). The O/C ratio is 0.56 calculated according to [48]. Fig. 3(b)-(e) present fitted and original spectra for Cys-GO/$Fe_3O_4$ composite's elements, namely, C1s, O1s, S2p and N1s, respectively. The presence of $Fe_3O_4$ binding energy peak at 530.42 eV implies for the successful decoration of Cys-GO.

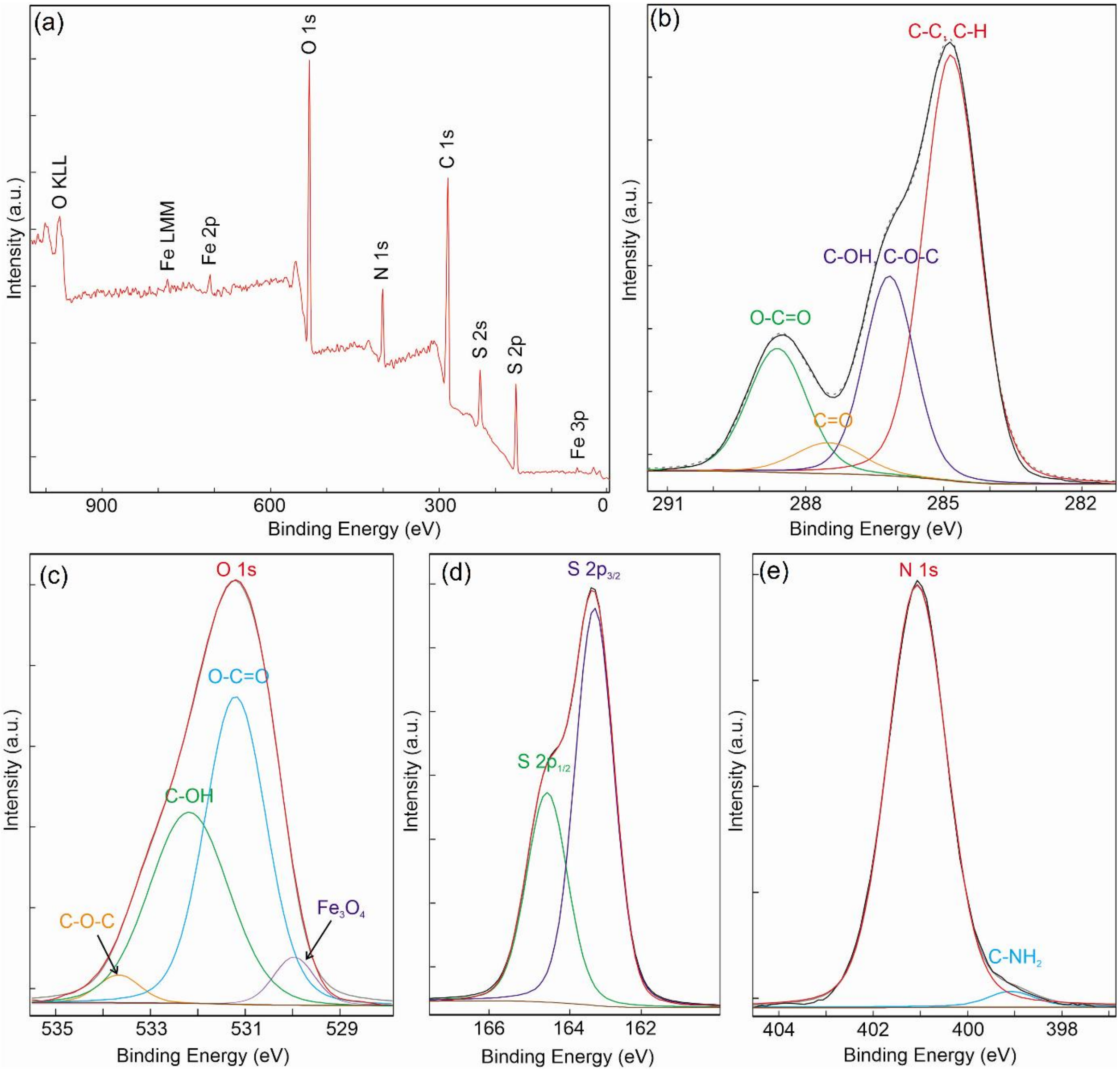

Fig. 3. XPS spectra of Cys-GO/$Fe_3O_4$ composite. (a) XPS survey scan spectra for Cys-GO/$Fe_3O_4$; (b); C1s (c) O1s; (d) S2p; (e) N1s peaks. Binding energies were referenced to the C1s main peak of carbon tape at 285eV.

Table 1 below represents binding energy peak positions for the Cys-GO/$Fe_3O_4$ composite.

Table 1. XPS binding energy peak positions for Cys-GO/$Fe_3O_4$.

<table>
<tr><th colspan="6">Cys-GO/$Fe_3O_4$</th></tr>
<tr><td colspan="6">Peak position (eV)</td></tr>
<tr><td rowspan="4">C1s</td><td>C-C, C-H</td><td>284.84</td><td rowspan="3">S$2p_{3/2}$</td><td colspan="2" rowspan="3">164.56</td></tr>
<tr><td>C-OH, C-O-C</td><td>286.18</td></tr>
<tr><td>C=O</td><td>287.50</td></tr>
<tr><td>O-C=O</td><td>288.61</td><td>S$2p_{1/2}$</td><td colspan="2">163.39</td></tr>
<tr><td rowspan="3">O1s</td><td>C-O-C</td><td>533.26</td><td>N1s</td><td>$NH_3^+$</td><td>401.07</td></tr>
<tr><td>C-OH</td><td>532.19</td><td rowspan="2"></td><td rowspan="2">$C-NH_2$</td><td rowspan="2">399.06</td></tr>
<tr><td>O-C=O</td><td>531.13</td></tr>
</table>

| | | | | | |
|---|---|---|---|---|---|
| | $Fe_3O_4$ | 530.42 | | | |

## 2.2 Magnetic field-induced orientation of $Fe_3O_4$-decorated Cys-GOLC

After synthesizing and characterizing aforementioned structures, we proceeded to study their orientational behaviour by applying a magnetic field. The uniform alignment of the LC molecules along the cell substrate and its controlling is crucial for LC-based device characteristics, response time, driving voltage, contrast ratio, and brightness [22]. For the alignment study of our materials the samples were examined by polarizing microscope. POM observations of the semidried samples indicating a random orientation of GOLC, GOLC/$Fe_3O_4$, and Cys-GOLC/$Fe_3O_4$ composites without presence of magnetic field (Fig. 4(a, c, e)). After applying a 0.3 T magnetic field perpendicular to the observation direction, the samples exhibited a distinct structural behaviour, namely aligned parallel to a magnetic field due to the magnetic field-induced orientation of nano- and microsheets (Fig. 4(b, d, f)) [37]. The noteworthy aspect here is that the orientation time, namely the full alignment of the samples under the applied magnetic field took 1 hour for GOLC, 30 seconds for GOLC/$Fe_3O_4$, and only 10 seconds for Cys-GOLC/$Fe_3O_4$, all under the same magnetic field strength [31]. Next, we investigated the effect of the magnetic orientation of our structures by UV-Vis spectroscopy method. An aqueous dispersions of above-mentioned materials were filled in a 1 mm thick quartz cuvette (40×10×1 mm). First, the sedimentation of our samples was investigated to exclude the presence of the latter in our system and ensure that the observed effects or changes are really the consequence of the influence of the magnetic field (see Fig. S2(a) in Supplementary Materials). Subsequently, a 0.3 T magnetic field was applied to the samples at room temperature. The magnetic field direction was perpendicular to the cuvette's surface. External field-dependent transmission spectra allow for the inference of the materials' reorientation. Subsequently, it was revealed that for three types of samples different times are needed to be subjected (Fig. 2(b-d) in Supplementary Materials) as was shown above through POM images. The decrease in the transmission spectra of GO after applying a magnetic field could be due to the alignment of GO sheets leading to increased optical density, scattering, or absorption, all of which reduce the amount of transmitted light. On the contrary, increase in transmission could be due to the MNP alignment in an applied magnetic field, which minimizes light absorption. For the Cys-GOLC/$Fe_3O_4$ the transmission spectrum might not change in the applied magnetic field because of the fast magnetic response saturation, or the material's fast optimal alignment, which was shown by POM observations in the main text. These measurements provide qualitative insights into the correlation between the applied magnetic field and the orientation of the GOLC systems [49].

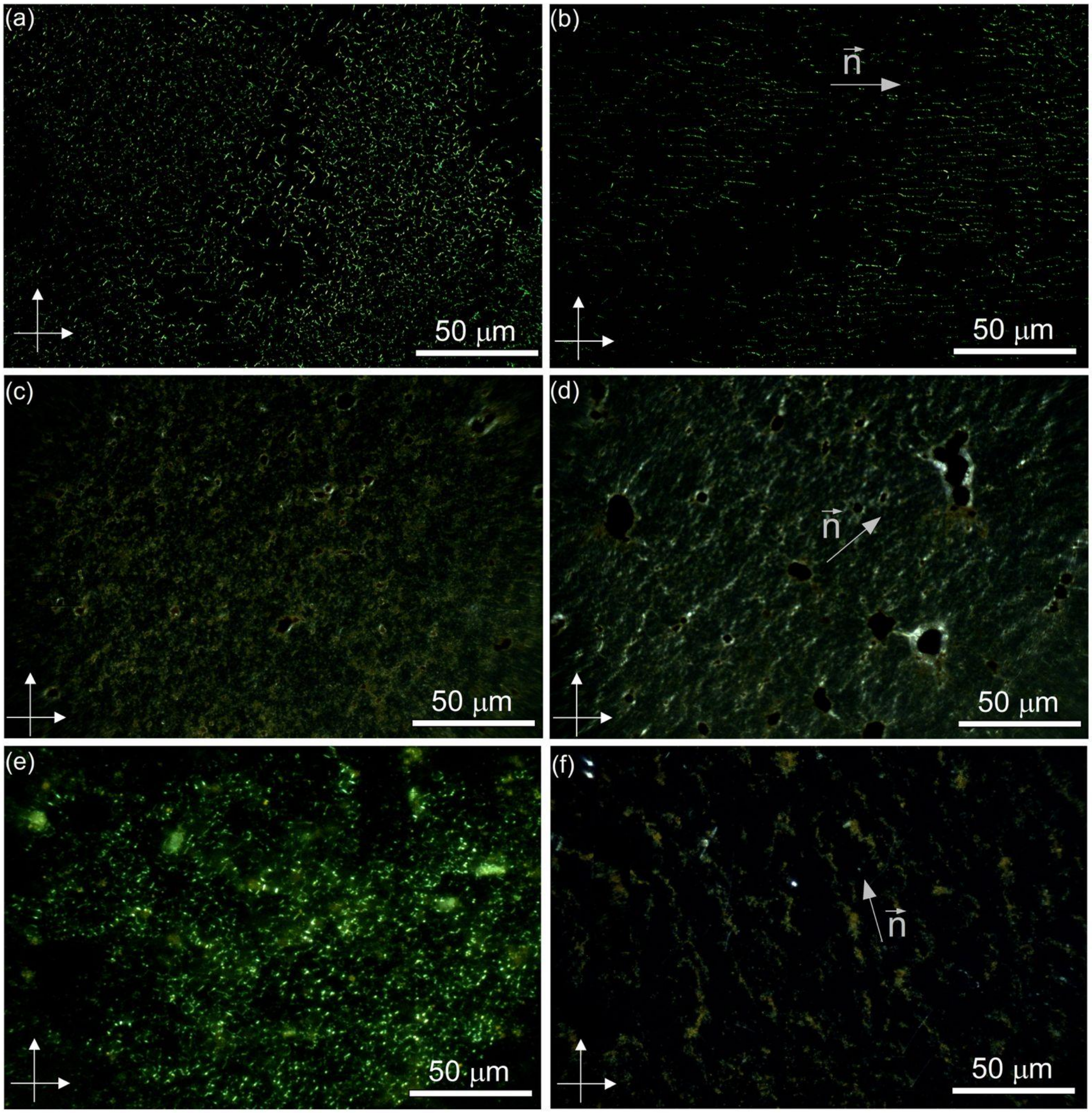

Fig. 4. POM micrographs of the samples between crossed polarizers: (a, c, e) without applying a magnetic field and (b, d, f) with the application of a 0.3 T magnetic field perpendicular to the observation direction for the GOLC, GOLC/$Fe_3O_4$ and Cys-GOLC/$Fe_3O_4$ systems, respectively.

Further, DLS measurements were also done to clarify relaxation behaviour of our systems. In this method the light scattered from a sample is detected at one well-defined scattering angle in a time-resolved way. The scattering angle during our measuments was automatically determined for each sample, based on its transmittance, which is continuously monitored. The time for each run (measurement time) was 10 s, and the measurements continued until 60 runs were completed. All measurements were performed at room temperature.

In the frame of this paper we present the results of photon correlation spectroscopy, which quantitatively describes signal fluctuations with the required time resolution. In correlation spectroscopy, a normalized autocorrelation function $g^{(2)}$ is calculated from the detected light-scattering intensity $I_{det}$ over the time *t*:

$$g^{(2)}(\tau) = \frac{\langle I_{det}(t) \cdot I_{det}(t+\tau) \rangle}{\langle I_{det}(t) \rangle^2} \quad (1)$$

The autocorrelation function $g^{(2)}$ of the light-scattering intensity represents the mean correlation between two measurements signals with time separation $\tau$. Normally, the reduced autocorrelation function $g^{(2)} - 1$ is calculated. The autocorrelation function decays exponentially with increasing $\tau$. The rate of this decay reflects the average speed of Brownian motion of the particles [50]. To estimate the relaxation rate, we performed a fitting of the intensity autocorrelation function using a double exponential model

$$y(t) = a_1 \exp\left(-\frac{t}{\tau_1}\right) + a_2 \exp\left(-\frac{t}{\tau_2}\right) \quad (2),$$

where $a_1$ and $a_2$ are amplitudes of the two components with $a_1 + a_2 = 1$ in normalized cases. $\tau_1$ and $\tau_2$ are time constants, namely the relaxation time which corresponds to the faster and slower decay. This fitting procedure enabled the determination of the relaxation time from the experimental autocorrelation function (Fig.4) [51, 52]. The relationship between the relaxation time ($\tau$) and the relaxation rate ($\upsilon$) can be approximated as $\upsilon = 1/\tau$.

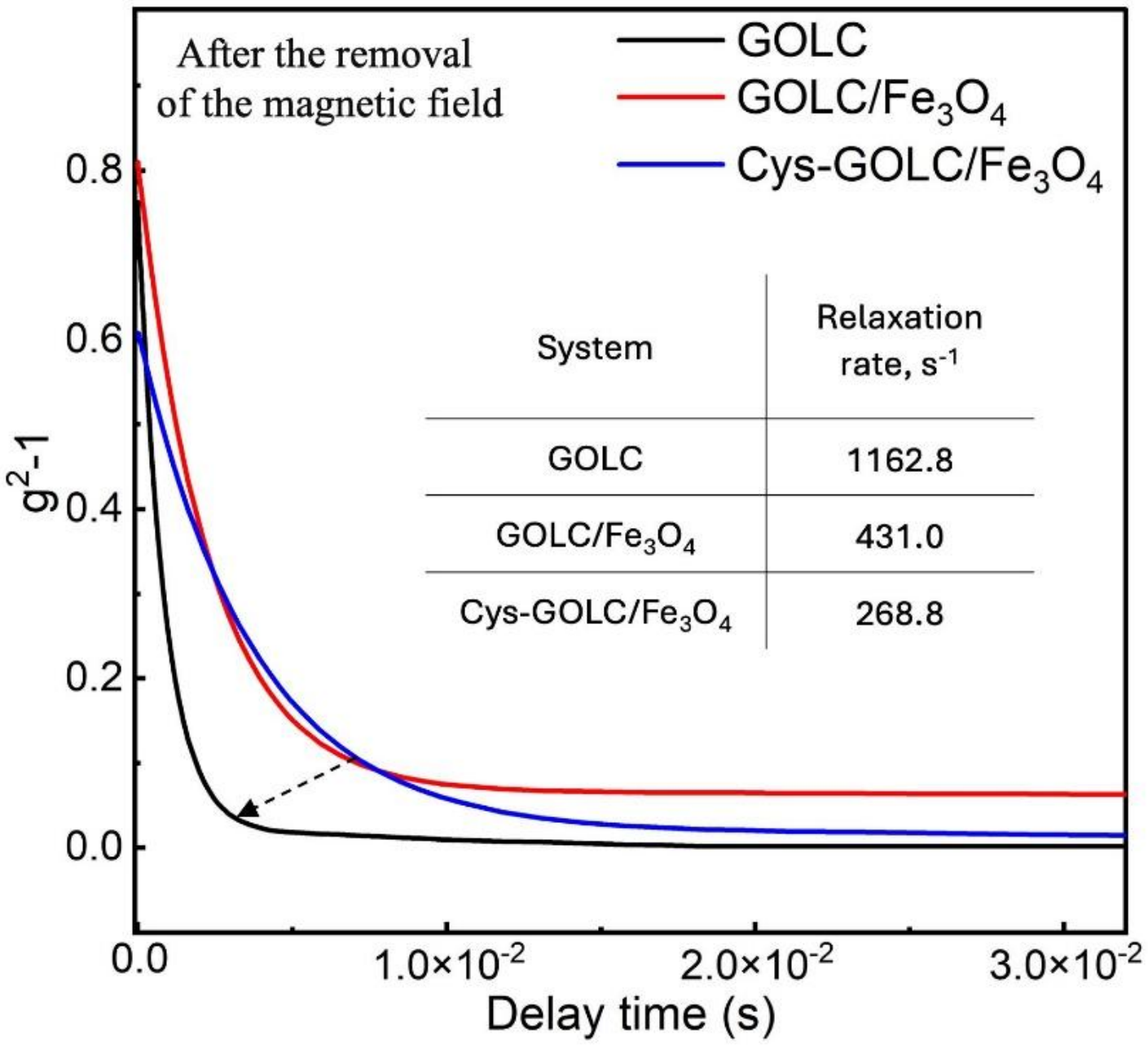

Fig. 5. Intensity autocorrelation function depending on time for GOLC, GOLC/$Fe_3O_4$, and Cys-GOLC/$Fe_3O_4$ systems after the removal of a 0.3 T magnetic field with relaxation rates. Laser light with a wavelength of 658 nm from a single-frequency laser diode, providing 40 mW of power, was used.

The fastest decay rate corresponds to the GOLC sample, while the slowest decay rate is observed for the Cys-GOLC/$Fe_3O_4$ sample.

Relaxation rate for our systems is presented on Fig.5. It is obvious that relaxation rate is lower for the Cys-GOLC/$Fe_3O_4$ composite. Therefore, on the one hand we claim that Cys decreases the reorientation/alignment time, and on the other hand we show that the relaxation time increases, thus proving the dual effect of Cys on our system. This dual effect could arise from the complex interplay between Cys molecules and the GOLCs, including interactions at the molecular level, surface modifications, and changes in the structural properties of GOLC. Additionally, factors such as concentration of Cys and $Fe_3O_4$ nanoparticles, as well as external stimuli parameters (e.g., magnetic field strength) would all influence the observed effects on the reorientation and relaxation time.

Our study on the dual effect of Cys on the orientation and relaxation of Cys-GOLCs/$Fe_3O_4$ in a magnetic field expands upon previous research exploring the magnetic alignment of graphene flakes. The magnetite-GO orientation under the external magnetic field is confirmed by the fading of the optical effect within 5 seconds after turning off the magnet [31]. This effect is observed in fields of 290 mT and 76 mT, with distinguishable light patterns even at a field strength of 10-13 mT, demonstrating a proportional relationship between the optical effect and the applied magnetic field. Besides, consistent with the findings in [49], we observe that the orientation of graphene flakes in a magnetic field leads to changes in optical properties, as the flakes align with the applied magnetic field. Unlike pure GO flakes, which required several hours to align due to their weak magnetism, our functionalized GOLC/$Fe_3O_4$ system aligns in just seconds, in line with the results presented in [15, 21]. This improvement is attributed to the stronger magnetic properties of $Fe_3O_4$ nanoparticles, which, when coupled with GOLC, enable rapid reorientation and alignment under the magnetic field. This coupling mechanism also facilitates the switchable photonic properties discussed in [29].

Extensive ordered patterns of GO on solid substrates, created by depositing aqueous dispersions through controlled drop evaporation is fast-developing field with huge potential for a wide range of innovative technologies [53]. Here, we show a simple method of large-scale micropatterning of GOLC systems and its controlling by external magnetic field. Fig. 6 (a, b) shows knit- and net-like patterns of GO created through colloidal-based drop plotting deposition. While the structure changes after applying a magnetic field is presented in Fig. 6(c). Strip-line patterns can be created in GOLC/$Fe_3O_4$ due to the localized distortion caused by the $Fe_3O_4$ nanoparticles within the structure (Fig. 6(d)). The emerged patterns depend on the size and concentration of GO flakes, as well as on the dewetting front of the drops. Micropatterning is also feasible for the Cys-GOLC/$Fe_3O_4$ system, since Cys acts as a reducing agent, enabling the patterned GOLCs to provide electrical conductivity and allowing precise control over the material's microscale structure.

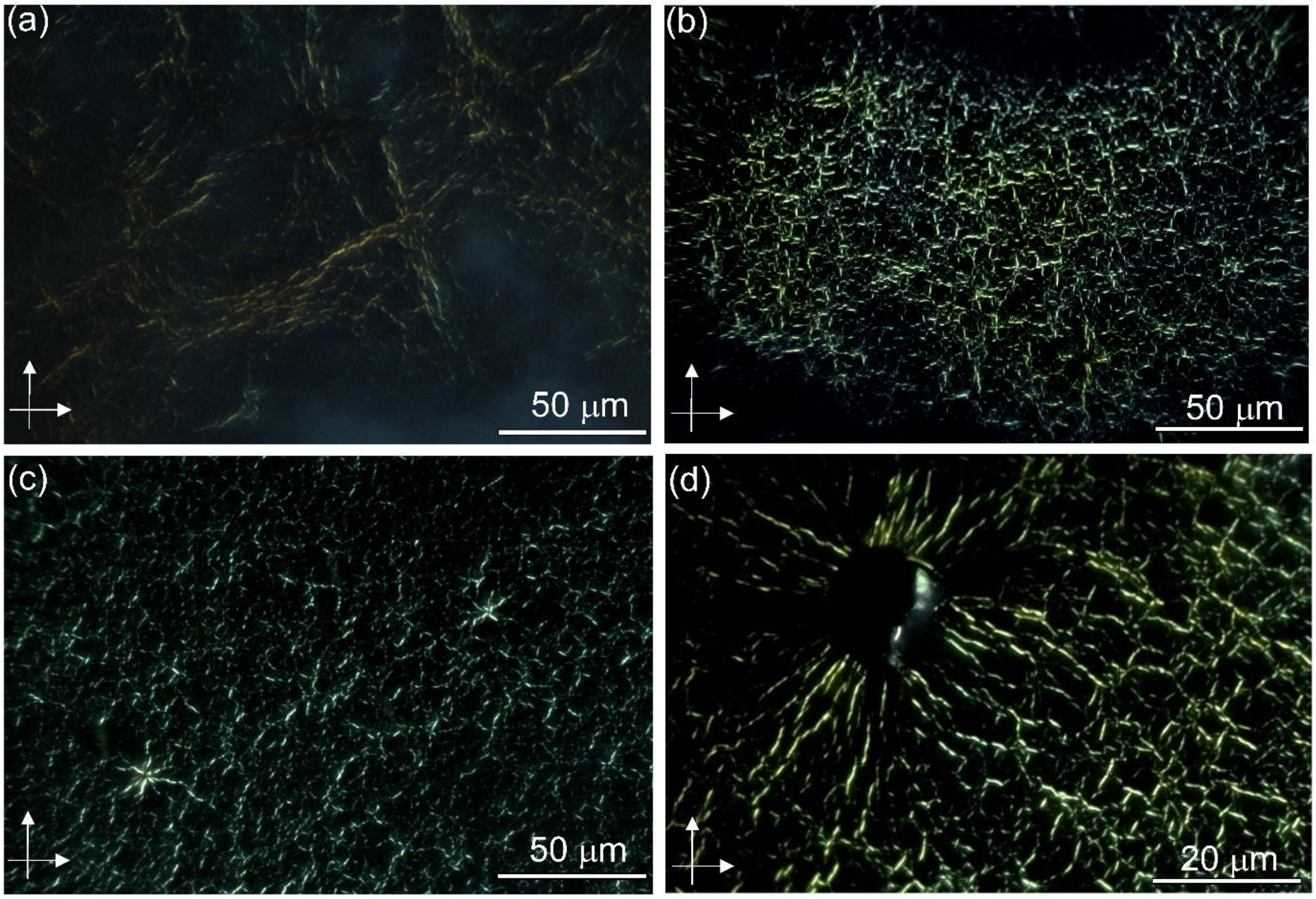

Fig. 6. Patterned GOLC samples. Drop dried and patterned (a) knit-like and (b) net-like GOLC without magnetic field; (c) repatterned net-like GOLC after the applying of a magnetic field; (d) repatterning caused by the distortion of the GOLC matrix due to the presence of $Fe_3O_4$ nanoparticles in the structure.

## Conclusion

In this study, the LC phase of electrochemically synthesized and Cys-functionalized GO was obtained. Further this structure was decorated by $Fe_3O_4$ MNPs synthesized through solution-combustion method. Comprehensive characterization was done to reveal this composite's structural and spectral properties. Additionally, a thorough comparison of the director behavior of GOLC, GOLC/$Fe_3O_4$, and Cys-GOLC/$Fe_3O_4$ was conducted under the applied magnetic field, utilizing detailed POM, DLS, and UV-Vis Spectroscopy methods. The findings demonstrate that Cys has dual-effect on the magnetic field-induced alignment and relaxation time of GOLC systems, namely decrease in reorientation time (from 1 hour to 10 s) and at the same time increase in relaxation time (an order of magnitude higher) after magnetic field influence. This can be observed due to various factors such as concentration and sizes of the constituent components, interaction mechanisms, and parameters of the experiment conditions. While fast response, *i.e.*, rapid changes in alignment, is essential for enhancing the efficiency of devices like biosensors, medical imaging systems, and drug delivery platforms, extended relaxation time improves contrast, resolution, and signal stability, thereby increasing the reliability of medical diagnostics and the effectiveness of treatment monitoring. Moreover, the creation and controlling of micropatterns in the drying drops of GOLC were demonstrated using a magnetic field. Micropatterns offer several advantages, such as enhanced light trapping and absorption, increased surface area, allowing for more charge carrier generation and improved charge collection, reduced reflection losses, tunable optical properties and mechanical stability. Thus, the controlled magneto-responsive behavior of these structures enabling a wide range of applications in fields such as sensing, drug delivery, environmental monitoring, and advanced anti-counterfeiting material design.